\documentstyle[axodraw]{article}
\begin{document}

\title{Quantum Algorithms: Database Search and its Variations}
\author{Apoorva Patel\\
{CHEP and SERC, Indian Institute of Science, Bangalore-560012}\\
{(E-mail: adpatel@cts.iisc.ernet.in)}}
\date{\today}
\maketitle

\centerline{\bf Abstract}
\begin{center}
\parbox{15truecm}{
The driving force in the pursuit for quantum computation is the exciting
possibility that quantum algorithms can be more efficient than their
classical analogues. Research on the subject has unraveled several aspects
of how that can happen. Clever quantum algorithms have been discovered in
recent years, although not systematically, and the field remains under
active investigation. This article is an introduction to the quantum
database search algorithm. Its extension to the quantum spatial search
algorithm is also described.}
\end{center}

\section{Quantum Computation}

Any physical system---with some initial state, some final state, and some
interaction in between---is a candidate for an information processing device,
i.e. a computer. One only needs to construct a suitable map between the
physical properties of the system and the desired abstract mathematical
variables. The initial state becomes the input, the final state becomes
the output, and the interaction provides appropriate logic operations.
Most of the development in theoretical computer science has been in the
framework of ``particle-like" discrete digital systems. The growth in
semiconductor technology has been so explosive---doubling the number of
transistors on a chip every 18-24 months according to Moore's law---that
many choices made in constructing the theoretical framework of computer
science (see for example, Ref.\cite{neumann}) were almost forgotten.
Computer architecture became essentially synonymous with digital electronic
circuits implementing Boolean operations, pushing aside other competing
models. It is known that ``wave-like" analogue computation can also be
carried out (e.g. using RLC circuits), but that has not been explored as
intensively. Even though a specific operation may be easier to implement
in digital mode than in analogue mode, or vice versa, considerations of
computational complexity are essentially the same for digital and analogue
algorithms. The choice between the two is therefore left to criteria for
hardware stability. Discrete variables are then a clear favourite---they
allow a degree of precision, by implementation of error correction
procedures, that continuous variables cannot provide.

The situation has changed with the advent of quantum computation. First
came the realisation that with shrinking size of its elementary components,
sooner or later, the computer technology will inevitably encounter the
dynamics of the atomic scale \cite{feynman}. The laws that apply at the
atomic scale are those of quantum mechanics and not those of electrical
circuits. The computational framework needs reanalysis, because quantum
objects display both ``particle-like" and ``wave-like" behaviour---the
discrete eigenstates that form the Hilbert space basis as well as the
superposition principle that allows for simultaneous existence of multiple
components. In what way would this combination alter the axioms of the
classical information theory? Is there much more to information theory
than just Boolean logic? Many investigations in quantum information theory
are addressed to such questions.

The next step was automatic. Clearly, with both ``particle-like" and
``wave-like" behaviour at their disposal, quantum algorithms can only
improve up on their classical counterparts. Note that the inputs and
the outputs of all the computational problems we investigate are always
classical (or are uniquely mapped to classical states); at the most a
quantum computer may solve a problem by a simpler non-classical algorithmic
route compared to the classical one. We clearly understand that the concept
of what is computable and what is not does not change in going from
classical to quantum computation. The scaling rules characterising how
efficiently a problem can be solved are altered, however, and the important
question is by how much. Explorations using several toy examples have
demonstrated that the improvement provided by a quantum solution, relative
to the corresponding Boolean logic solution, depends on the problem. The
extraordinary feature is that in certain cases the difference is large
enough to challenge the conventional complexity classification of
computational problems.

The key ingredient for the superiority of quantum algorithms over the
Boolean ones happens to be the parallelism provided by the superposition
of quantum states. A typical quantum algorithmic strategy exploits it by
superposing an exponentially large number of quantum components using
only polynomial resources. For example, with $n$ qubits and $n$ rotations,
one can create a uniform superposition of $2^n$ quantum components:
\begin{equation}
|0\rangle^{\otimes n} \longrightarrow
\left(\frac{|0\rangle+|1\rangle}{\sqrt{2}}\right)^{\otimes n}
= 2^{-n/2} \sum_{i=0}^{2^n-1} |i\rangle ~.
\end{equation}
A quantum instruction applied to a superposed state processes all the
superposed components together, in a manner analogous to the classical
SIMD (single instruction multiple data) paradigm, but without the need
for any extra parallel computation resources. Thus a single run of a
quantum algorithm can take $2^n$ superposed inputs to $2^n$ superposed
outputs. The caveat is that a projective measurement of the output can
give only one of the output components while erasing all the rest, not
unlike the situation that one can listen to only a single radio or
television programme at a time from a superposition of a large number
of available signals. The huge advantage inherent in superposition is
therefore useful in those computational problems, where many different
inputs need to be processed by the same instructions but only one
specific property of the possible outputs is desired at the end. A clear
criterion for identifying such problems has not been found, although
pattern recognition problems are good candidates.

When the full exponential advantage provided by superposition can be
used, the complexity classification of the quantum solution becomes
exponentially better than the classical one. Indeed, a lot of research
effort has been directed towards discovering problems whose classical
solution is in class NP (non-deterministic polynomial), while the quantum
solution would be in class P (polynomial). But even in cases where the
advantage provided by superposition can only partially be used, the
improvement in the scaling rules of the solutions can be substantial
enough for real life applications.

Major boost to the subject of quantum computation was provided by the
discovery of efficient quantum solutions to two problems of practical
interest. One is Shor's algorithm for factoring large integers \cite{shor},
which is based on an exponentially faster quantum Fourier transform
solution to the period finding problem compared to the classical fast
Fourier transform one. The other is Grover's algorithm for finding an
object in an unsorted database \cite{grover}, which provides a square
root speed-up by cleverly using interference to enhance the amplitude
at the desired location while suppressing the amplitudes at all other
locations. Grover's algorithm applies to unstructured problems, and
assumes so little in terms of either the properties of the problem or
the instructions to be executed, that it is generally believed that it
will be a popular application on any quantum computer in one form or
the other. The following sections provide an introduction to this
wonderful algorithm and some of its variations.

\section{The Database Search Problem}

A database is a collection of items. These items possess certain properties,
which can be used to distinguish them from each other. The problem of database
search is to find an item with the desired properties from the collection.
An efficient search process is the one which can locate the desired item in
the database as quickly as possible. In practice, efficiency of the search
process becomes more and more important as the database size grows.

To understand the optimisation of the search process, it is instructive to
first look at a familiar game played by school children. In the game, two
teams compete to find the names of famous persons. One team selects the name
of a famous person and gives it to the referee. The other team has to discover
this name by asking a set of questions to the first team. The game is made
interesting by the condition that the first team only provides ``yes or no"
answers to the questions, i.e. the minimal amount of information---one bit---is
released in response to every question. The two teams take turns choosing the
names and asking the questions. After several rounds, the team which succeeds
in discovering the names with a smaller number of questions is the winner.

The players quickly learn that specific questions such as ``Is the person
Richard Feynman?" are inefficient. They fail most of the time, and when they
fail one does not learn much about who the person is. Efficient questions
are of the type ``Is the person a man or a woman?", whence there is a
substantial reduction in the number of possible names for the next step no
matter what the answer to the question is. The best questions are the ones
which reduce the number of possible names for the next step by a factor of
two. This factor of two reduction is of course a consequence of the fact
that the answers provided to the questions are binary.

A computer scientist would describe this game as ``database search". The
list of possible names forms the database, the questions are referred to as
queries, and the answers provided by the other team are called oracles. The
winning strategy is the one that finds the hidden name with least number of
oracle consultations. Binary search is the optimal classical algorithm, which
finds the desired item in a database of size $N$ using $Q=\log_2 N$ binary
queries. This algorithm can be used when the items in the database are sorted
according to some order, so that at every step the items corresponding to
``yes" answer can be easily separated from the items corresponding to ``no".
The ordering can be implicit (e.g. the team asking "man or woman" question
already knows how to separate the two possibilities), or it can be explicit
(e.g. the manner in which the words are arranged in a dictionary).

Sorting the items means arranging them in an ordered sequence according to
their property values. Though property values along the real line are
sufficient for establishing the sequence, the values are taken to be distinct,
and without loss of generality they can be replaced by integer labels.
The integer labels can be easily digitised, i.e. written as a string of
letters belonging to a finite alphabet. When the alphabet has $a$ letters,
a string of $n$ letters can label $N=a^n$ items. (If the number of items in
the database is not a power of $a$, then the database can be padded up with
extra labels to make $N=a^n$. We also assume for simplicity that there are
no duplicate entries.) In digital computers, this finite alphabet has the
smallest size, i.e. $a=2$, and the letters are called bits.

Sorting facilitates subsequent searching of a database, by factorisation of
the search process. Though digitisation does not change the order of the items
in the sequence, it simplifies the search process. To locate an item in the
database with known properties, one does not look for the complete string of
letters in one go, but sequentially inspects only one letter of the label at
a time. With digitisation, the individual search steps have to distinguish
amongst only a limited number of possibilities, and the maximum simplification
of the search steps occurs when the alphabet has the smallest size, i.e. $a=2$.
Sorting requires significant effort---$O(N \log_2 N)$ operations for a database
of size $N$ \cite{knuth}.
But once a database is sorted, all subsequent searches to locate any item in
it require only $O(\log_2 N)$ queries.

The advantage of sorting becomes obvious when one looks at the process of
finding the desired item in an unsorted database, e.g. finding the name of
the person with a given telephone number using a telephone directory.
There is no classical quick search algorithm available---random pickings
are as good as any other selection scheme. All one can do is to go through
all the items one by one, asking the binary question ``Is this the item
that I want or not?"; no further reduction in search effort is possible.
On the average, one requires $\langle Q\rangle=N$ binary queries to locate
the desired item in an unsorted database, since each query has a success
probability of $1/N$. The number of queries can be reduced to $\langle
Q\rangle=(N+1)/2$, provided that the search process has a memory so that
an item rejected once is not picked up again for inspection. Now we can
understand that it is the large change in the number of queries, from
$O(N)$ to $O(\log_2 N)$, that makes the laborious process of sorting
worthwhile, to be carried out once and for all.

\section{Grover's Quantum Solution}

Can one improve upon this classical analysis? The answer is yes, if we give
up some implicit assumptions in the methodology. A crucial assumption in the
classical analysis has been that one can inspect only one item of the database
(or one letter of the label) at a time. This assumption can be bypassed in a
quantum algorithm, where one can work with a superposition of states and
``inspect" many---even all---items with a single query. Lov Grover exploited
this superposition feature of quantum dynamics, and discovered the optimal
quantum database search algorithm \cite{grover}.
The quantum algorithm is far more interesting for an unsorted database than
for a sorted one, and we look at that in detail first.

Quantum states are unit vectors in a Hilbert space (linear vector space with
complex coefficients). They evolve in time by unitary transformations. These
properties are totally different from classical states and their evolution,
and they form the basis of the subject of quantum computation. Of course, to
make contact with our problems defined in classical language, we require a
mapping between classical and quantum states. That is achieved by identifying
the orthogonal basis vectors of the Hilbert space with the set of distinct
classical states. The complex components of a general vector in the Hilbert
space can vary continuously, and they are called amplitudes of the quantum
state. States with more than one non-zero amplitudes are superposed states.
Quantum algorithms evolve the amplitudes from some initial values to some
final values, by a sequence of unitary transformations. The measurement
probability of obtaining a particular classical result from a quantum state
is given by the absolute value squared of the corresponding amplitude.

The quantum database search algorithm works in an $N$-dimensional Hilbert
space, whose basis vectors are identified with the individual items.
It takes an initial state whose amplitudes are uniformly distributed
over all the items, to a final state where all but one amplitudes vanish.
Following Dirac's notation, let $\{|i\rangle\}$ be the set of basis vectors,
$|s\rangle$ be the initial uniform superposition state, and $|t\rangle$ be
the final state corresponding to the desired item. Then
\begin{equation}
|\langle i|s \rangle| = 1/\sqrt{N} ~,~~ \langle i|t \rangle = \delta_{it} ~~.
\end{equation}
At any stage of the algorithm, the amplitude corresponding to a particular
item can be obtained by projecting the quantum state along the corresponding
basis vector, using the operator $P_i = |i\rangle\langle i|$.

Since superposition of amplitudes is commutative, the quantum database can
be taken to be unsorted without any loss of generality. The optimal solution
to the quantum search problem is based on two properties: (i) the shortest
path between two points on the unitary sphere is the geodesic great circle,
and (ii) the largest step one can take in a given direction, consistent with
unitarity, is the reflection operation. The available reflection operators are:
\begin{equation}
U_t = 1 - 2|t\rangle\langle t| ~,~~ U_s = 1 - 2|s\rangle\langle s| ~~.
\end{equation}
(A projection operator satisfies $P^2 = P$, which makes $(1-2P)$ a reflection
operator, i.e. $(1-2P)^2 = 1$.) The operator $U_t$ distinguishes between the
desired state and the rest. It flips the sign of the amplitude in the desired
state, and is the binary query or the quantum oracle. The operator $-U_s$
treats all items on an equal footing; it flips the sign of the amplitudes
relative to the initial uniform state. Note that applying $P_s$ to any state
$|k\rangle$ averages all its component amplitudes, and so the operation $-U_s$
is often called ``reflection in the average".

Grover's algorithm is the discrete Trotter's formula with these two operators,
which locates the desired item in the database with $Q$ queries,
\begin{equation}
{\cal G}^Q |s\rangle \equiv (-U_sU_t)^Q |s\rangle = |t\rangle ~~.
\end{equation}
This iterative procedure is readily evaluated to yield the relation
\begin{equation}
(2Q+1) \sin^{-1} (1/\sqrt{N}) = \pi/2 ~~.
\label{QNformula}
\end{equation}
For a given $N$, the solution for $Q$ satisfying Eq.(\ref{QNformula}) may not
be an integer. This means that the algorithm will have to stop without the
final state being exactly $|t\rangle$. There will remain a small admixture
of other amplitudes in the output, implying an error in the search process.
The size of the unwanted admixture is determined by how close one can get to
$\pi/2$ on the r.h.s. of Eq.(\ref{QNformula}). That restricts the error rate
to be smaller than $1/N$. Apart from this, the algorithm is completely
deterministic.

\subsection{Geometric Interpretation}

The algorithm has a simple geometric structure. It follows from the
definitions above that the evolving quantum state always remains in the 
2-dimensional subspace spanned by the vectors $|s\rangle$ and $|t\rangle$.
The state therefore proceeds from its initial value $|s\rangle$ towards
its target value $|t\rangle$ along the geodesic connecting the two.

Let the initial angle between $|s\rangle$ and $|t_\perp\rangle$ be $\theta$.
Fig.1 depicts how a state $|k\rangle$ in the $|t\rangle-|s\rangle$ subspace
transforms under the action of the two operators, $U_t$ and $-U_s$. When the
angle between $|k\rangle$ and $|t_\perp\rangle$ is $\phi$, one application
of $U_t$ changes it to $-\phi$, and one application of $(-U_sU_t)$ changes
that to $\phi+2\theta$. Thus the quantum state rotates at a uniform rate of
$2\theta$ per query in the direction $|t_\perp\rangle \rightarrow |t\rangle$.
It is easy to figure out that the eigenvalues of the rotation operator
${\cal G}$ are $e^{\pm2i\theta}$, and the corresponding eigenvectors are
$(1,\pm i)/\sqrt{2}$ in the $(|t\rangle, |t_\perp\rangle)$ basis.

\begin{figure}[h]
{
\begin{picture}(400,140)
  \thicklines
  \put(160,10){\vector(1,0){120}}
  \put(285,10){\makebox(0,0)[bl]{$|t\rangle$}}
  \put(160,10){\vector(0,1){120}}
  \put(155,135){\makebox(0,0)[bl]{$|t_\perp\rangle$}}
  \put(160,10){\vector(1,2){53.7}}
  \put(210,125){\makebox(0,0)[bl]{$|k\rangle$}}
  \put(160,10){\vector(1,4){29.1}}
  \put(185,132){\makebox(0,0)[bl]{$|s\rangle$}}
  \put(160,10){\vector(-1,2){53.7}}
  \put(100,125){\makebox(0,0)[bl]{$U_t|k\rangle$}}
  \CArc(160,10)(28,76,90)
  \CArc(160,10)(30,76,90)  \put(162,45){\makebox(0,0)[bl]{$\theta$}}
  \CArc(160,10)(45,63,90)  \put(176,58){\makebox(0,0)[bl]{$\phi$}}
  \CArc(160,10)(40,90,117) \put(148,55){\makebox(0,0)[bl]{$\phi$}}
  \put(160,10){\vector(4,3){96}}
  \put(240,90){\makebox(0,0)[bl]{$-U_sU_t|k\rangle$}}
  \CArc(160,10)(54,76,117)
  \CArc(160,10)(56,76,117) \put(135,70){\makebox(0,0)[bl]{$\phi+\theta$}}
  \CArc(160,10)(60,37,76)
  \CArc(160,10)(62,37,76)  \put(200,65){\makebox(0,0)[bl]{$\phi+\theta$}}
  \CArc(160,10)(90,37,63)  \put(220,85){\makebox(0,0)[bl]{$2\theta$}}
\end{picture}
}
\caption{The steps of the quantum database search algorithm as rotations
in the plane formed by the states $|s\rangle$ and $|t\rangle$. The angle
$\theta$ is defined by the relation $\langle t|s\rangle=\sin\theta$.}
\vspace*{-2mm}
\label{fig:rotation}
\end{figure}
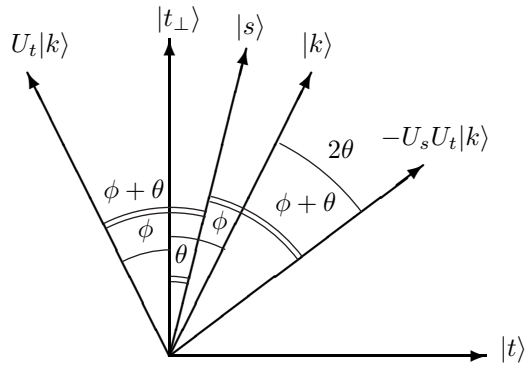

Clearly, successive applications of ${\cal G}$ to $|s\rangle$ change
the angle from $\theta$ to $3\theta, 5\theta, 7\theta, 9\theta, \ldots$.
We obtain the result in Eq.(\ref{QNformula}), noting that the initial
angle is determined by
\begin{equation}
\sin\theta = \langle t|s\rangle = 1/\sqrt{N} ~~,
\end{equation}
and the total desired rotation after $Q$ queries is $\pi/2 - \theta$.
In general, $\pi/2$ is not an odd-integral multiple of $\theta$. Then
the best approximation to the desired state is obtained by running the
algorithm till the quantum state gets within angle $\theta$ of $|t\rangle$.
Note that to obtain the optimal solution, one has to know precisely when
to stop the algorithm, i.e. the value of $Q$ (or equivalently $N$) should
be known apriori. In case $N$ is known only in order of magnitude and not
precisely, one can stop the algorithm randomly after $O(\sqrt{N})$ queries.
Since the iterative algorithm rotates the state at a uniform rate in the
$|t\rangle-|s\rangle$ subspace, the success probability over a number of
random attempts is $\langle\sin^2\phi\rangle=1/2$. That is still much
larger than the initial value $1/N$ and only a factor of two worse than
the best result.

This picture makes it clear that the algorithm is optimal, i.e. it proceeds
along the shortest geodesic path and takes the largest possible steps
(operators other than reflection operators would produce rotation angle
less than $2\theta$). No other algorithm, classical or quantum, can locate
the desired object in an unsorted database with a fewer number of queries
\cite{BBBV,zalka}.

\subsection{The Smallest Solution}

It is also interesting to visualise the evolution of amplitudes in case of
the smallest solution to Eq.(\ref{QNformula}). This solution is exact and
corresponds to locating the desired item in a database of four items with
a single query. It is quite a contrast to the classical case, where a single
binary query can distinguish only two items. Fig.2 illustrates how the
evolution operators change various amplitudes in the 4-dimensional Hilbert
space. The algorithm amplifies the desired amplitude and eliminates the
unwanted amplitudes, by a clever interference amongst them. Note that this
is possible only when the amplitudes have both positive and negative
values---a characteristic wave feature that is absent in Boolean logic
operations. Thus the algorithm illustrates a quantum amplification process,
where the desired amplitude grows at the expense of the rest. This internal
rearrangement of amplitudes is unusual, in stark contrast to conventional
amplifiers that require the help of an external energy source.

\begin{figure}[h]
{
\setlength{\unitlength}{1mm}
\begin{picture}(120,75)
  \thicklines
  \put( 5,65){\makebox(0,0)[bl]{(1)}}
  \put(12,65){\line(1,0){32}}
\put(13,73){\line(1,0){2}} \put(17,73){\line(1,0){2}} \put(21,73){\line(1,0){2}}
\put(25,73){\line(1,0){2}} \put(29,73){\line(1,0){2}} \put(33,73){\line(1,0){2}}
\put(37,73){\line(1,0){2}} \put(41,73){\line(1,0){2}}
  \put(45,65){\makebox(0,0)[bl]{0}} \put(45,73){\makebox(0,0)[bl]{0.5}}
  \put(16,65){\line(0,1){8}} \put(24,65){\line(0,1){8}}
  \put(32,65){\line(0,1){8}} \put(40,65){\line(0,1){8}}
  \put(55,70){\makebox(0,0)[bl]{Uniform distribution}}
  \put(100,70){\makebox(0,0)[bl]{Equilibrium}}
  \put(100,66){\makebox(0,0)[bl]{configuration}}
  \put(28,62){\vector(0,-1){8}}
  \put(30,58){\makebox(0,0)[bl]{$U_t$}}
  \put(55,58){\makebox(0,0)[bl]{Quantum oracle}}
  \put(100,58){\makebox(0,0)[bl]{Binary query}}
  \put( 5,45){\makebox(0,0)[bl]{(2)}}
  \put(12,45){\line(1,0){32}}
\put(13,49){\line(1,0){2}} \put(17,49){\line(1,0){2}} \put(21,49){\line(1,0){2}}
\put(25,49){\line(1,0){2}} \put(29,49){\line(1,0){2}} \put(33,49){\line(1,0){2}}
\put(37,49){\line(1,0){2}} \put(41,49){\line(1,0){2}}
  \put(45,45){\makebox(0,0)[bl]{0}} \put(45,49){\makebox(0,0)[bl]{0.25}}
  \put(16,45){\line(0,-1){8}} \put(24,45){\line(0,1){8}}
  \put(32,45){\line(0,1){8}} \put(40,45){\line(0,1){8}}
  \put(55,47){\makebox(0,0)[bl]{Amplitude of desired}}
  \put(55,43){\makebox(0,0)[bl]{state flipped in sign}}
  \put(100,47){\makebox(0,0)[bl]{Sudden}}
  \put(100,43){\makebox(0,0)[bl]{perturbation}}
  \put(28,34){\vector(0,-1){8}}
  \put(30,30){\makebox(0,0)[bl]{$-U_s$}}
  \put(55,30){\makebox(0,0)[bl]{Reflection about average}}
  \put(100,30){\makebox(0,0)[bl]{Over-relaxation}}
  \put( 5,15){\makebox(0,0)[bl]{(3)}}
  \put(12,15){\line(1,0){32}}
\put(13,19){\line(1,0){2}} \put(17,19){\line(1,0){2}} \put(21,19){\line(1,0){2}}
\put(25,19){\line(1,0){2}} \put(29,19){\line(1,0){2}} \put(33,19){\line(1,0){2}}
\put(37,19){\line(1,0){2}} \put(41,19){\line(1,0){2}}
  \put(45,15){\makebox(0,0)[bl]{0}} \put(45,19){\makebox(0,0)[bl]{0.25}}
  \put(16,15){\line(0,1){16}}
  \put(24,15){\circle*{1}} \put(32,15){\circle*{1}} \put(40,15){\circle*{1}}
  \put(55,20){\makebox(0,0)[bl]{Desired state reached}}
  \put(100,20){\makebox(0,0)[bl]{Opposite end}}
  \put(100,16){\makebox(0,0)[bl]{of oscillation}}
  \put( 5,5){\makebox(0,0)[bl]{(4)}}
  \put(12,5){\makebox(0,0)[bl]{Projection}}
  \put(55,5){\makebox(0,0)[bl]{Algorithm is stopped}}
  \put(100,5){\makebox(0,0)[bl]{Measurement}}
\end{picture}
}
\vspace*{-1mm}
\caption{The steps of the quantum database search algorithm for the simplest
case of 4 items, when the first item is desired by the oracle. The left column
depicts the amplitudes along the 4 basis vectors, with the dashed lines showing
their average values. The middle column describes the algorithmic steps, and
the right column mentions their physical implementation in the wave language.}
\vspace*{-2mm}
\label{fig:database}
\end{figure}
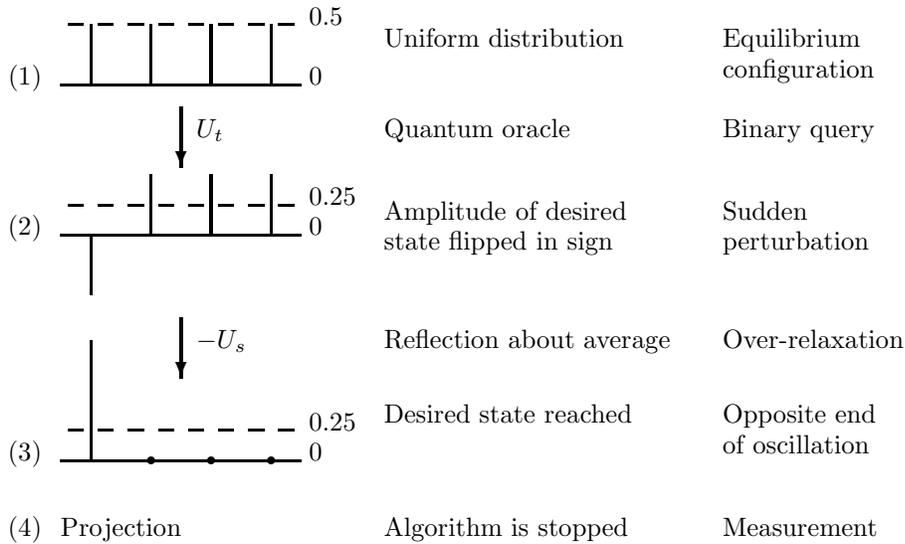

\subsection{Salient Features}

There are several noteworthy features of this algorithm:
\begin{itemize}
\itemsep=0pt
\item   The iterative steps of the algorithm can be viewed as the discretised
evolution of the quantum state in the Hilbert space, governed by a Hamiltonian
containing two terms, $|t\rangle\langle t|$ and $|s\rangle\langle s|$. The
former represents a potential energy attracting the state towards $|t\rangle$,
while the latter represents a kinetic energy isotropically diffusing the state
throughout the Hilbert space \cite{trotter}. The alternate application of
$U_t$ and $U_s$ in the discretised evolution steps is then reminiscent of
Trotter's formula, generated by exponentiating the terms in the Hamiltonian
when constructing the the transfer matrix from a discretised path integral.
\item   Asymptotically, $Q = \pi\sqrt{N}/4$. The best that the classical
algorithms can do is to random walk through all the possibilities, and that
produces $Q = O(N)$ as mentioned above. With the use of superposition of all
possibilities at the start, the quantum algorithm performs a directed walk
along the geodesic to the final result and achieves a square-root speed-up.
Rather rarely one comes across such truly optimal algorithms.
\item   It is easy to see that the result in Eq.(\ref{QNformula}) depends
only on $|\langle t|s\rangle|$ and not on $\langle t|s\rangle$. The phases of
various components of $|s\rangle$ can therefore be arbitrary (e.g. they can
have the symmetry of bosons, fermions or even anyons). The starting point of
the algorithm can therefore be generalised from the state with all amplitudes
equal to the state with all probabilities equal.
\item   It is straightforward to convert the algorithm to the situation
where there are $M$ marked items in the database instead of just one.
Then $|t\rangle$ is replaced by $\sum_{k=1}^M |t_k\rangle/\sqrt{M}$, and
the algorithm yields one of the marked items after $O(\sqrt{N/M})$ queries.
\item   In case the quantum database permits factorised queries, i.e.
inspection of one letter of the item label at a time, the algorithm can
be speeded up just as in the classical situation. Mere digitisation of item
labels is sufficient for efficient search, however, sorting becomes redundant
with the quantum superposition of states. Since a single quantum query can
distinguish one of four items, the factorised quantum search algorithm can
locate the desired item with $\log_4 N$ queries. This is a factor of two
improvement on the best classical algorithm applied to a sorted database
\cite{assembly}.
\item   Applicability of the algorithm has been extended to the much wider
context of amplitude amplification \cite{bhmt}. In this case, the state
$|s\rangle$ is replaced by $V|s\rangle$, where $V$ is any unitary operator.
The overlap $\langle t|s\rangle$ then gets replaced by $\langle t|V|s\rangle$,
and the operator $U_s$ gets replaced by $V U_s V^\dagger$. This generalised
algorithm amplifies the initial probability $|V_{ts}|^2$ towards unity,
and can be used to improve the situation where $V$ is another algorithm with
a small success probability.
\item   The algorithm does not require the full power of quantum dynamics.
It only needs superposition of states and interference of amplitudes. So it
can very well be implemented using any system---not necessarily quantum---that
obeys the superposition principle. For example, explicit examples using
classical waves in the form of coupled oscillators have been constructed
\cite{wavesrch1,wavesrch2}. In such mechanical systems, the role of the
uniform superposition state $|s\rangle$ is played by the centre-of-mass mode.
Fig.2 also describes the individual steps of the algorithm in the wave
language. The number of queries required in the wave scenario and in the
quantum case is the same. The difference is that to represent $N$ items of
the database one requires $N$ wave modes but only $\log_2 N$ qubits.
\item   The quantum algorithm for searching an unsorted database does not
assume any specific structure or pattern as far as the input data are
concerned. This makes it highly versatile with broad utility. The algorithm
has been incorporated, either in full or in part, in the quantum solutions
of a variety of problems. Some of such applications are: square-root speed-up
of solutions to NP-complete problems, finding mean and median of statistical
distributions, locating minimum of a function, quantum counting, deciphering
cryptographic codes, and so on. Details can be found in the literature
\cite{nielsen,archive}.
\item   The solutions of Eq.(\ref{QNformula}) for $Q=1,2,3$, viz. $N=4,10.5,
20.2$ respectively, are of special significance for the number of building
blocks involved in genetic information processing \cite{quant_gc,genetic_info}.
Living organisms generally do not have access to desired biomolecules in a
ready-made form. They first break down the ingested food into small building
blocks, and then assemble the pieces in a precise manner to synthesise the
desired biomolecules. The desired pieces are selected by pattern recognition
binary oracles (i.e. whether correct molecular bonds will be formed or not)
provided by the master template available in DNA. Such assembly processes
follow the unsorted database search algorithm with factorised queries.

The DNA alphabet has four letters identified by a single base-pairing, and
the amino acid alphabet has twenty letters identified by three consecutive
base-pairings. Both these languages are universal, i.e. the same all the way
from viruses and bacteria to human beings. The numbers involved fit Grover's
algorithm, and not the powers of two associated with the classical binary
search. The binary alphabet is certainly simpler, and hence likely to appear
earlier during evolution. So did the living organisms evolve more complicated
genetic languages because of the advantage offered by Grover's optimal
algorithm? That is a tantalising possibility which can explain the unanswered
mystery at the origin of life---why do genetic languages have the structure
that they do?---not as an accident of history but as the optimal solution to
the implemented task.
\end{itemize}

\section{The Spatial Search Problem}

It can happen that the items belonging to a physical database are spread
over distinct locations instead of being in one place, and one still needs
to find the desired item among them. When there is a restriction that one
can proceed from any location to only its neighbours, while inspecting
the unsorted items, we have the spatial search problem. This problem is
conveniently represented using a graph, with the vertices denoting the
locations of items and the edges labeling the connectivity of neighbours.
Boolean algorithms for this problem are $O(N)$, since they can do no
better than inspect one location after another until reaching the desired
item. Grover's algorithm illustrates that one can do better in search
problems by working with a superposition of states, but it uses the
global operator $U_s$, with the property $\langle j|U_s|i\rangle=-2/N$
for any $i\ne j$. That is tantamount to moving between any two locations,
not just neighbours, with equal ease. To solve spatial search problems,
therefore, we need to investigate algorithms where the operator $U_s$ is
replaced by a local operator that connects only neighbouring locations.
The scaling behaviour of such algorithms is expected to depend on the
database size $N$ as well as the connectivity of the graph.

Working with a regular graph (i.e. the same number of neighbours for
every vertex) keeps the problem tractable. In fact, most investigations
of spatial search problems have been carried out on hypercubic lattices,
where the dimensionality $d$ specifies the connectivity of the graph.
In this geometry, $N=L^d$ and each vertex has $2d$ neighbours. Grover's
algorithm can then be looked up on as the $d\rightarrow\infty$ limit,
i.e. the mean-field-theory limit of statistical mechanics where any
vertex is a neighbour of any other one. Furthermore, by varying both
$N$ and $d$ independently, we can develop a broad picture of how the
dimension of the database influences the scaling behaviour of the spatial
search problem, and how various limits are approached.

Complexity of quantum spatial search on a hypercubic lattice obeys two
simple lower bounds. One arises from the fact that while the marked vertex
could be anywhere on the lattice, a local step can move from any vertex to
only its neighbours. Since the marked vertex cannot be located without
reaching it, the worst case scenario requires $\Omega(dL)$ steps. This
bound weakens with increasing $d$, and is the strongest in one dimension
where a quantum algorithm cannot improve on the $O(N)$ classical algorithm.
The other bound follows from the fact that spatial search cannot outperform
Grover's optimal algorithm, which has no restriction on movement. Quantum
spatial search therefore must require $\Omega(\sqrt{N})$ oracle queries.
This bound is independent of $d$, and stronger than the first bound for
$d>2$. Combined together, the two bounds make the complexity of quantum
spatial search $\Omega(d N^{1/d},\sqrt{N})$. Note that the two bounds are
of the same magnitude, $\Omega(\sqrt{N})$, for the critical case of $d=2$.

The way these lower bounds arise in spatial search illustrates an interplay
of two distinct physical principles, special relativity (or no faster than
light signalling) and unitarity. The locality constraint on movement is the
result of the finite propagation speed in relativity. On the other hand,
the optimality of Grover's algorithm is a consequence of unitarity
\cite{BBBV,zalka}. It is known that the two principles are compatible,
although just barely, and so the best spatial search algorithms should
arise in a framework that respects both the principles, i.e. relativistic
quantum mechanics.

The concepts of critical phenomena in statistical mechanics turn out to be
useful in understanding the scaling behaviour of spatial search algorithms
in different dimensions: (i) Universality of scaling suggests that the
scaling exponents depend on the dimensionality of the database but not
on further details of graph connectivity. (ii) Increasing the number of
neighbours for a vertex makes the local movement restriction less relevant
and takes the system toward its mean-field-theory limit. (Note that the
maximum value of $d$ is $\log_2 N$ for finite $N$.) (iii) Interplay of
different dynamical features produces logarithmic correction factors in 
critical dimensions due to infrared divergences. Indeed, spatial search
algorithms in $d=2$ are slowed down by extra logarithmic factors, whose
suppression by clever algorithm design (including use of additional
parameters) is an interesting exercise.

We point out that preparation of the unbiased initial state for the quantum
spatial search problem, i.e. preparation of the translationally invariant
uniform superposition state $|s\rangle=\sum_x|\vec{x}\rangle/\sqrt{N}$,
does not add to the complexity of the problem. $|s\rangle$ can easily be
prepared using local directed steps. For instance, one can start at the
origin, then step by step transfer the amplitude to the next vertex along
an axis, and achieve an amplitude $L^{-1/2}$ at all the vertices on the
axis after $L$ steps. Thereafter, repeating the procedure for each remaining
coordinate direction produces the state $|s\rangle$ after $dL$ steps in
total. This cost is smaller than the lower bounds on quantum spatial search
mentioned above.

\section{Quantum Random Walk Based Solution}

We follow Grover's quantum algorithmic strategy for spatial search as well,
i.e. construct a Hamiltonian evolution where the kinetic part of the
Hamiltonian diffuses the amplitude distribution all over the lattice and
the potential part of the Hamiltonian attracts the amplitude distribution
toward the marked vertex \cite{trotter}. The change is that we have to
replace the global evolution operator $U_s$ (corresponding to the kinetic
part of the Hamiltonian) by a local movement operator $W$, and then optimise
$W$ to achieve the fastest diffusion \cite{hypsrch1}. The typical discrete
method for exploring an unstructured space, with the constraint of local
movements, is random walk. In the ``particle" form, that is a diffusion
process which spreads according to $distance \propto \sqrt{time}$. On the
other hand, a coherent ``wave" form can spread faster according to $distance
\propto time$, and we certainly want to use that in a quantum algorithm.

Classical random walks evolve the given probability distribution in a
non-deterministic manner at every time step. They have been used to
tackle a wide variety of graph theory problems, usually with a local
and translationally symmetric evolution rule. We use a quantum version
of this process, i.e. quantum random walks \cite{qrw}. It provides a
unitary evolution of the quantum amplitude distribution, such that the
amplitude at each vertex gets redistributed over itself and its neighbours
at every time step. Quantum random walks are deterministic, unlike
classical random walks, with quantum superposition allowing simultaneous
exploration of multiple possibilities. Several quantum algorithms have
used them as important ingredients, and an introductory overview can be
found in Ref.\cite{qrwrev}.

The spatial propagation modes of a quantum random walk are characterized
by their wave vectors $\vec{k}$. Quantum diffusion depends on the energy
of these modes according to
\begin{equation}
U(\vec{k},t) = \exp(-iE(\vec{k})t) ~~.
\end{equation}
The lowest energy mode, $\vec{k}=0$, corresponding to a uniform distribution,
is an eigenstate of the diffusion operator and does not propagate. The
slowest propagating modes are the ones with the smallest nonzero $|\vec{k}|$.
The commonly used diffusion operator is the Laplacian, e.g. it appears
in the non-relativistic Schr\"odinger equation. That gives $E(\vec{k})
\propto |\vec{k}|^2$ corresponding to the $distance \propto \sqrt{time}$
spread. On the other hand, the relativistic massless Dirac operator has
the dispersion relation $E(\vec{k})\propto|\vec{k}|$ corresponding to the
$distance \propto time$ spread. Its quadratically faster diffusion of the
slowest modes makes it better suited to quantum spatial search algorithms
\cite{gridsrch1,gridsrch2}.

\subsection{Spatial Search with the Dirac Operator}

An automatic consequence of the Dirac operator is the appearance of an
internal degree of freedom corresponding to spin, whereby the quantum
state becomes a multi-component spinor. These spinor components can guide
the diffusion process, and be interpreted as the states of a coin
\cite{gridsrch1,gridsrch2}. While this is the only possibility for the
continuum theory, another option is available for a lattice theory,
i.e. the staggered fermion formalism \cite{staggered}. In this approach,
the spinor degrees of freedom are spread in coordinate space over an
elementary hypercube, instead of being in an internal space. Using it
for quantum search reduces the total Hilbert space dimension by $2^d$
and eliminates the coin toss instruction.

The Dirac Hamiltonian in the continuum theory is well-known, and we need to
construct a discrete time evolution operator corresponding to its kinetic
part. Note that even when the Hamiltonian $H$ is local, the associated
unitary evolution operator $U=\exp(-iH\tau)$ may connect arbitrarily far
apart vertices. To make $U$ also local, we split $H$ in to block-diagonal
Hermitian parts and then exponentiate each part separately \cite{qwalk1}.
For a bipartite lattice, partitioning of $H$ in to two parts (which we label
``odd" and ``even") is sufficient for this purpose:
\begin{equation}
H = H_o + H_e ~.
\end{equation}
Each part then contains all the vertices, but only half of the links
attached to each vertex. Each link is associated with a term in $H$
providing propagation along it, and appears in only one of the two parts.
On a hypercubic lattice, $H$ gets divided in to a set of non-overlapping
blocks of size $2^d \times 2^d$ (each block corresponds to an elementary
hypercube) that can be exactly exponentiated. After this bipartite
decomposition, we let the quantum random walk evolve according to
\begin{equation}
\psi(\vec{x};t) = W^t \psi(\vec{x};0) ~,~~
W = U_e U_o = e^{-iH_e\tau} e^{-iH_o\tau} ~.
\end{equation}
Each block of the unitary matrices $U_{o(e)}$ mixes amplitudes of vertices
belonging to a single elementary hypercube, and the amplitude distribution
spreads in time because the two alternating matrices do not commute. Note
that $W$ does not perform evolution according to $H$ exactly. Instead,
$W=\exp(-iH\tau) + O(\tau^2)$. Still, the truncation is such that $W$ is
exactly unitary, i.e. $W=\exp(-i\tilde{H}\tau)$ for some $\tilde{H}$.

The search algorithm is completed by alternating the kinetic evolution step
with the potential evolution step. Just as in Grover's algorithm, the largest
potential evolution step corresponds to making the evolution phase maximally
different from $1$. That is the binary oracle reflection operator,
\begin{equation}
R = I - 2 |\vec{0}\rangle\langle\vec{0}| ~,
\end{equation}
when the marked vertex is chosen to be the origin. Thus the spatial search
algorithm evolves the amplitude distribution according to
\begin{equation}
\psi(\vec{x};t_1,t_2) = [W^{t_1} R]^{t_2} \psi(\vec{x};0,0) ~.
\label{evolsearch}
\end{equation}
Here $t_2$ is the number of oracle queries, and $t_1$ is the number of
quantum random walk steps between queries. Both should be minimised to
find the the quickest solution to the spatial search problem.

Grover's algorithm is designed to evolve the quantum state in the
two-dimensional subspace spanned by the initial and the desired states.
It strides along the geodesic arc from $|s\rangle$ to $\vec{0}\rangle$
perfectly, and reaches the marked vertex with the maximum probability
$P_{\rm max}=1$. That does not hold for spatial search. The amplitude
distribution evolving with a local operator $W$ (instead of the global
operator $U_s$) does not remain fully confined to the two-dimensional
subspace. The maximum probability to reach the marked vertex is therefore
reduced, $P_{\rm max}<1$. The remedy is to augment the algorithm by the
amplitude amplification procedure \cite{bhmt}, and reach the marked vertex
with $\Theta(1)$ probability. The complexity of the algorithm then increases
to the effective number of oracle queries, $t_2/\sqrt{P_{\rm max}}$.

To obtain the best results, one can tune parameters such that $W^{t_1}$
approximates $-U_s$ as closely as possible, i.e. maximise the overlap
$Tr(-W^{t_1}U_s)$. This condition can be recast as the minimisation of
$\langle\vec{0}|W^{t_1}|\vec{0}\rangle$, which is the quantum random walk
amplitude to return to the starting point. In practice, this has to be
performed numerically, depending on the parameters appearing in $W^{t_1}$.

\subsection{Numerical Results}

We have performed numerical simulations of the quantum spatial search
algorithm in various dimensions \cite{dgt2srch,deq2srch,fracsrch}, as per
the theoretical considerations described above, and observed the following:
\begin{itemize}
\itemsep=0pt
\item   For large values of $t_1$, the amplitude distribution diffuses
too much out of the two-dimensional subspace, and the algorithm fails
to reach the marked vertex with any meaningful probability. The best
results for complexity as well as total computational cost are obtained
with $t_1=2$ or $3$.
\item   The eigenvalues of $U_s$ are $\pm1$, while those of $W$ are
spread around the unit circle. We find that, with optimal tuning of the
parameters, $W^{t_1}$ approximates $-U_s$ in the average sense, i.e.
$W^{t_1}$ is a reflection operator for the average eigenvalue of the
associated $\tilde{H}^2$. Appearance of the average eigenvalue in this
result indicates that all spatial modes contribute to the search process
with roughly equal strength.
\item   For $d>2$, the optimised quantum spatial search algorithm has the
same scaling behaviour as Grover's algorithm. As $N\rightarrow\infty$,
$P_{\rm max}$ approaches a constant and $t_2$ is proportional to $\sqrt{N}$.
The locality restriction on movement does not matter much, and the slow down
with respect to Grover's optimal algorithm is only in the scaling prefactor.
Looking at it differently, the square-root speed-up provided by relativity
(through change in dispersion relation) is perfectly compatible with that
provided by unitarity.
\item   With increasing $d$, the prefactor in complexity scaling approaches
its optimal value $\pi/4$. Moreover, we can get pretty close to the optimal
value for all $d>2$. In particular, the scaling prefactor for $d=3$ exceeds
$\pi/4$ by about 25\%, and that for $d=7$ by about 10\%. The approach to the
asymptotic scaling behaviour as a function of $d$ is illustrated in Fig.3,
where $N$ is changed by keeping $L$ fixed and changing $d$. It can also be
seen that for fixed $N$, it is best to implement the algorithm using the
smallest $L$ and hence the largest $d$.
\end{itemize}

\begin{figure}[h]
\epsfxsize=10cm
\centerline{\epsfbox{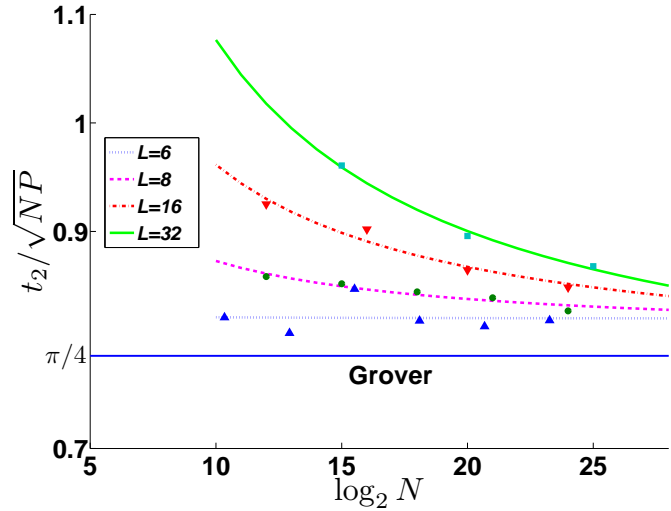}}
\caption{Effective number of oracle queries as a function of the database
size for $d=3$ to $9$ and different lattice sizes. The points are the data
for $t_1=3$, and the curves are the fits $t_2/\sqrt{NP}=a+b/d$. Also shown
is the limiting value corresponding to Grover's optimal algorithm.}
\end{figure}

\begin{itemize}
\itemsep=0pt
\item   For $d\le 2$, the evolution operator arising from the massless
Dirac operator is infrared divergent (as $\int d^dk/k^2$ in the continuum
formulation). That slows down spatial search algorithms, altering the
behaviour of both $P_{\rm max}$ and $t_2$ compared to their optimal scaling
forms, by logarithmic factors in the critical dimension $d=2$ and by power
law factors in $d<2$. In particular, the infrared divergence causes the
amplitude distribution to evolve too much out of the two-dimensional
subspace, and the maximum probability to reach the marked vertex plummets
to zero as the database size increases. One way to regulate the infrared
divergence is to introduce a non-zero mass term in the Dirac operator.
That regulates the infrared divergence through $k^2 \rightarrow k^2+m^2$,
but also slows down the diffusion process. For small enough $m$, the
diffusion speed (and hence $t_2$) may not change much, but substantial
change in the contribution of the $|\vec{k}| \le m$ modes can modify
the scaling of $P_{\rm max}$. Then an optimal value of $m$ can be obtained
by trading off the increase in $t_2$ against the increase in $P_{\rm max}$.
For a finite lattice, the lattice size acts as the infrared cutoff, and so
we expect the optimal value of $m$ to be a function of the database size $N$.
\item   Tulsi constructed an infrared regulated spatial search algorithm
possessing the above described properties \cite{tulsi}. In his algorithm
the $R$ and $W$ operations are controlled by an ancilla qubit such that the
quantum walk pauses (i.e. misses some $W$ steps) when it passes through the
marked vertex. This concentration of the quantum walk at the marked vertex
can also be looked up on as the appearance (only at the marked vertex) of
an effective mass in the propagator or a self-loop in the evolution graph.
The effect is controlled by a mixing parameter $\cos\delta$ ($\cos\delta=1$
means no pausing). Tulsi showed that for $\cos\delta=\Theta(1/\sqrt{\ln N})$,
the amplitude concentration at the marked vertex cuts down the amplitude
diffusion out of the two-dimensional subspace, and increases $P_{\rm max}$
to $\Theta(1)$ without affecting the scaling of $t_2$.
\item   The scaling behaviour of the infrared regulated quantum spatial
search algorithm in $d=2$ is illustrated in Fig.4. Our best result for
the complexity is $t_2/\sqrt{P} \approx 0.45 \sqrt{N\log_2 N}$.
\end{itemize}

\begin{figure}[h]
\epsfxsize=10cm
\centerline{\epsfbox{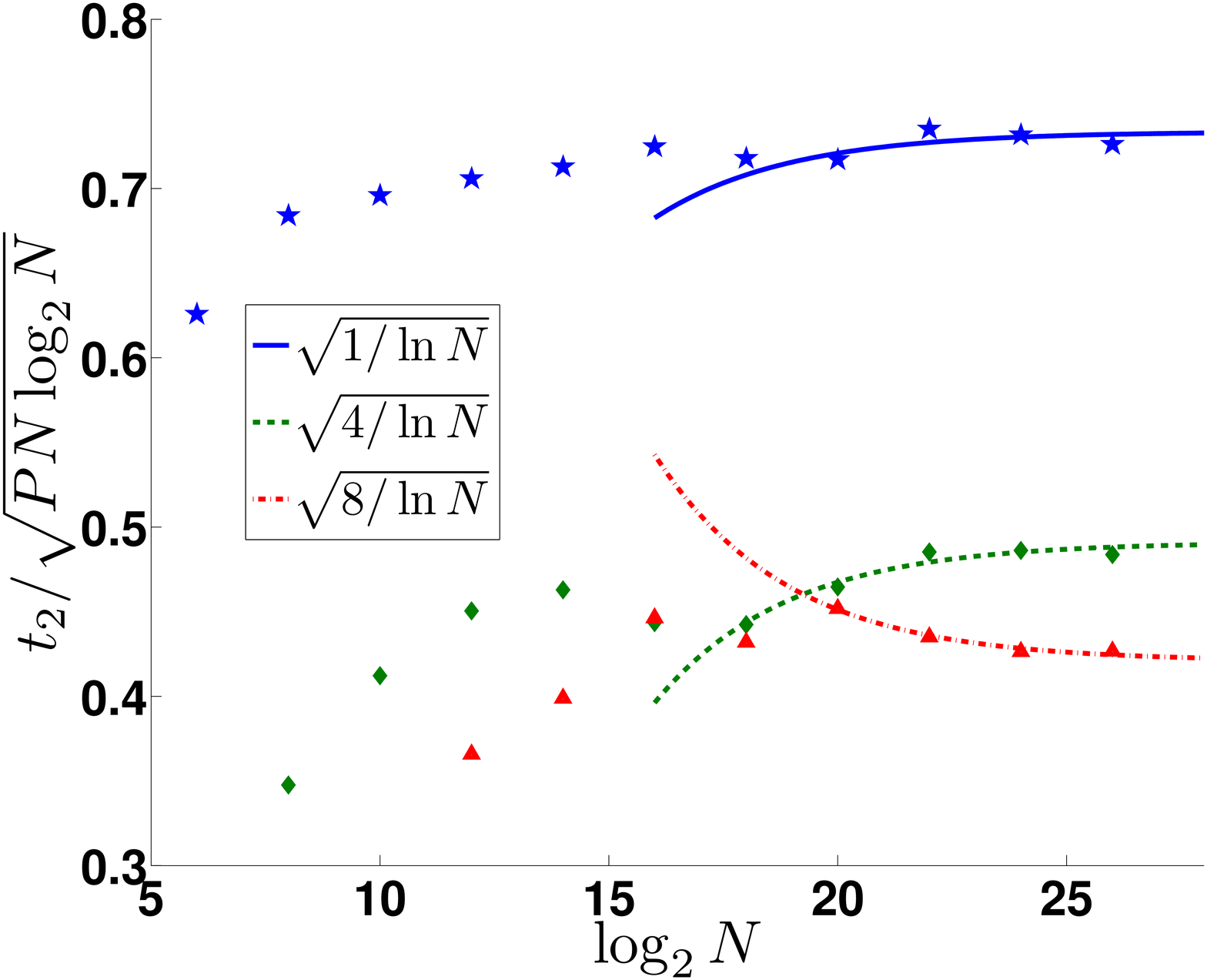}}
\caption{Effective number of oracle queries as a function of the database
size for $d=2$. The points are the data for $t_1=3$ and different values
of the ancilla control parameter $\cos\delta$. The curves are the fits
$t_2/\sqrt{PN\log_2 N}=a+b/L$.}
\end{figure}

\begin{itemize}
\itemsep=0pt
\item   Scaling behaviour of spatial search algorithms in non-integer
dimensions can be studied using fractal lattices. That is useful for
understanding the complexity scaling bounds, particularly for $1<d<2$,
and testing the analogy with critical phenomena in statistical mechanics.
Fractal lattices do not have translational invariance or a definition of
the Dirac operator. Nevertheless, relativistic diffusion can be produced
on regular lattices using the flip-flop operator \cite{gridsrch1}.
Our numerical results for the infrared regulated algorithm on Sierpinski
gaskets show that the complexity does scale as $N^{1/d}$, where $d$ is
the spectral dimension (i.e. the dimension of the reciprocal lattice
modes $\vec{k}$) and not the fractal (i.e. Hausdorff) dimension.
\item   The spatial search algorithm with a trap at the marked vertex
has a potential application in understanding dynamics of photosynthesis.
Light harvesting complexes of photosynthetic organisms have energy transport
efficiency exceeding 95\%, and the exciton motion is experimentally found
to be quantum coherent for more than $400fs$ \cite{photosynth1,photosynth2}.
The complex consists of a reaction centre surrounded by light absorbing
pigment molecules in an antenna structure. The energy of the absorbed
photon is transported to the reaction centre in a wave motion, and has
to remain there long enough till it is used in chemical reactions. The
details of how this occurs without dissipation are still to be understood.
But a coupled system of wave modes, in a specific geometry and with a
trap at a particular location, can be a good model for understanding the
phenomenon as an optimised solution.
\end{itemize}

\section{Outlook}

All known quantum algorithms are constructed in the SIMD paradigm, with
the quantum parallelism superposing $N$ amplitudes using only $\log_2 N$
qubits. At its best, that can reduce the complexity of quantum algorithms
compared to classical ones by a factor of $N/\log_2 N$. Efficient
quantum algorithms also need factorisation of the unitary operations in
the $N-$dimensional Hilbert space in to smaller blocks. That can also
provide a speed-up of $O(N/\log_2 N)$ under the best circumstances.
(With full factorisation, $N$ terms can be produced by multiplying
$\log_2 N$ two-term factors.) Factorisation is however a classical
computational procedure, and the advantage provided by it may or may
not overlap with that provided by quantum parallelism. Efficiency of a
quantum algorithm therefore depends on how these two features interplay.

In case of Shor's solution to the period finding problem \cite{shor},
complete classical factorisation converts the Fourier transform operation
to FFT, yielding a speed-up of $O(N/\log_2 N)$. When FFT is converted to
QFT, the quantum parallelism provides another non-overlapping gain of
$O(N/\log_2 N)$. The result is a large separation between the classical
and the quantum complexity of the problem. In the database search problem,
gains of factorisation (i.e. sorting) and quantum parallelism overlap.
So one cannot obtain two speed-up factors of $O(N/\log_2 N)$---the quantum
factorised algorithm provides only a factor of two speed-up over the
classical factorised algorithm. In case of Grover's algorithm, the oracle
is not factorised. That limits the gain that can be extracted from quantum
parallelism, and only a square-root speed-up can be achieved. We do not
know any systematic procedure to identify the problems amenable to both
quantum parallelism and classical factorisation, and generic problems
would have neither. Still, problems that possess both these features form
an interesting hunting territory for efficient quantum algorithms, and we
must explore that to come up with new quantum algorithms.

\end{document}